\documentclass[twoside, journal]{IEEEtran}

\usepackage{epsfig}
\usepackage{graphicx}
\usepackage{amsmath}
\usepackage{amssymb}
\usepackage{float}
\usepackage{url}
\newtheorem{definition}{Definition}

\newtheorem{proposition}{Proposition}
\newtheorem{corollary}{Corollary}
\newtheorem{theorem}{Theorem}

\newtheorem{remark}{Remark}
\newtheorem{example}{Example}

\newcommand{\naturals}{\ensuremath{\mathbb{N}}}
\newcommand{\reals}{\ensuremath{\mathbb{R}}}

\newcommand{\pr}{\ensuremath{\mathbb{P}}}
\newcommand{\expectation}{\ensuremath{\mathbb{E}}}

\begin{document}
\markboth{2012 IEEE Information Theory Workshop, Lausanne,
Switzerland, September 4--7, 2012.}{I. SASON: On the Entropy of Sums of Bernoulli
Random Variables via the Chen-Stein Method}

\title{ \huge{On the Entropy of Sums of Bernoulli Random
Variables via the Chen-Stein Method}}

\author{\IEEEauthorblockN{Igal Sason\\
Department of Electrical Engineering\\
Technion - Israel Institute of Technology\\
Haifa 32000, Israel\\
\hspace*{-0.4cm} E-mail: sason@ee.technion.ac.il}}

\maketitle

\thispagestyle{empty}
\pagestyle{empty}

\begin{abstract}
This paper considers the entropy of the sum of
(possibly dependent and non-identically distributed) Bernoulli
random variables. Upper bounds on the error that follows from an
approximation of this entropy by the entropy of a Poisson
random variable with the same mean are derived. The derivation
of these bounds combines elements of information theory with the
Chen-Stein method for Poisson approximation.
The resulting bounds are easy to compute, and their applicability
is exemplified. This conference paper presents in
part the first half of the paper entitled
``An information-theoretic perspective of the Poisson approximation
via the Chen-Stein method'' (see: \url{http://arxiv.org/abs/1206.6811}).
A generalization of the bounds that considers the
accuracy of the Poisson approximation for the entropy of a sum of non-negative,
integer-valued and bounded random variables is introduced in the full paper.
It also derives lower bounds on the total variation distance, relative entropy
and other measures that are not considered in this conference paper.
\end{abstract}

\vspace*{-0.35cm}
\begin{keywords} Chen-Stein method, entropy, information theory,
Poisson approximation, total variation distance.
\end{keywords}

\vspace*{-0.3cm}
\section{Introduction}
\label{section: introduction}
Convergence to the Poisson distribution, for the number of occurrences of possibly
dependent events, naturally arises in various applications. Following the work of
Poisson, there has been considerable interest in how well the Poisson distribution
approximates the binomial distribution. This approximation was treated by a limit
theorem in \cite[Chapter~8]{Feller_book1950}, and later some non-asymptotic theoretical
results have studied the accuracy of this approximation.
The Poisson approximation and
later the compound Poisson approximation have been treated extensively in the probability
and statistics literature (see, e.g., \cite{ArratiaGG_AOP}--\cite{Chen_1975},
\cite{DasGupta_2008}--\cite{Feller_book1950},
\cite{Le Cam_1960}--\cite{Steele_1994} and references therein).

Among modern methods, the Chen-Stein method forms a powerful probabilistic tool that is used
to calculate error bounds when the Poisson approximation serves to assess the distribution of
a sum of (possibly dependent) Bernoulli random variables \cite{Chen_1975}. This method is
based on the simple property of the Poisson distribution where $Z \sim \text{Po}(\lambda)$
with $\lambda \in (0, \infty)$ if and only if
$ \lambda \, \expectation[f(Z+1)] - \expectation[Z \, f(Z)]= 0 $
for all bounded functions $f$ that are defined on $\naturals_0 \triangleq \{0, 1, \ldots \}$.
This method provides a rigorous analytical treatment, via error bounds, to the
case where $W$ has approximately a Poisson distribution $\text{Po}(\lambda)$ so it is
expected that $ \lambda \, \expectation[f(W+1)] - \expectation[W \, f(W)] \approx 0 $
for an arbitrary bounded function $f$ that is defined on $\naturals_0$. The 
reader is referred to some nice surveys on the
Chen-Stein method in \cite{ArratiaGG_Tutorial90}, \cite{BarbourHJ_book_1992},
\cite[Chapter~2]{BarbourC_book_2005}, \cite{Probability_Surveys_2005},
\cite[Chapter~2]{RossP_book07}, \cite{Ross_Tutorial11}.

\newpage
During the last decade, information-theoretic methods were exploited to establish
convergence to Poisson and compound Poisson limits in suitable paradigms. An
information-theoretic study of the convergence rate of the binomial-to-Poisson
distribution, in terms of the relative entropy between the binomial and Poisson
distributions, was provided in \cite{HarremoesR_2004}, and maximum entropy results
for the binomial, Poisson and compound Poisson distributions were studied in
\cite{Harremoes_2001}, \cite{Johnson_2007}, \cite{KarlinR_81}, \cite{Shepp_Olkin_1981},
\cite{Yu_IT08}, \cite{Yu_IT09_paper1} and \cite{Yu_IT09_paper2}. The law of small
numbers refers to the phenomenon that, for random variables $\{X_i\}_{i=1}^n$ on
$\naturals_0$, the sum $\sum_{i=1}^n X_i$ is approximately Poisson distributed with mean
$\lambda = \sum_{i=1}^n p_i$ as long as (qualitatively) the following conditions hold:
\begin{itemize}
\item $\pr(X_i = 0) \approx 1$, and $\pr(X_i = 1)$ is uniformly small,
\item $\pr(X_i > 1)$ is negligible as compared to $\pr(X_i=1)$,
\item $\{X_i\}_{i=1}^n$ are weakly dependent.
\end{itemize}
An information-theoretic study of the law of small numbers was provided in
\cite{KontoyiannisHJ_2005} via the derivation of upper bounds on the relative
entropy between the distribution of the sum of possibly dependent Bernoulli
random variables and the Poisson distribution with the same mean. An extension
of the law of small numbers to a thinning limit theorem for convolutions of
discrete distributions that are defined on $\naturals_0$ was introduced in
\cite{Thinning_IT2010} followed by an analysis of the convergence rate and
some non-asymptotic results. Further work in this direction was studied in
\cite{JohnsonY_IT10}, and the work in  \cite{BarbourJKM_EJP_2010}
provides an information-theoretic study for the problem of compound Poisson
approximation, which parallels the earlier study for the Poisson approximation
in \cite{KontoyiannisHJ_2005}. Nice surveys on this line of work are provided
in \cite[Chapter~7]{Johnson_2007}, \cite{Kontoyiannis_slides2006}, and
\cite[Chapter~2]{DasGupta_2008} surveys some commonly-used metrics between
probability measures with some pointers to the Poisson approximation.

This paper provides an information-theoretic study of Poisson approximation,
and it combines elements of information theory with the Chen-Stein method.
The novelty in this paper,
in comparison to previous related works, is related to the derivation
of upper bounds on the error that follows from an approximation of the
entropy of a sum of possibly dependent and non-identically distributed
Bernoulli random variables by the entropy of a Poisson random variable
with the same mean (see
Theorem~\ref{theorem: upper bound on the Poisson approximation of the entropy}
and some of its consequences in Section~\ref{section: Error bounds on the
entropy of the sum of Bernoulli random variables}).
The use of these new bounds is exemplified,
partially relying on interesting applications of the Chen-Stein method
from \cite{ArratiaGG_Tutorial90}.

\newpage
\section{Error Bounds on the Entropy of the Sum of Bernoulli
Random Variables}
\label{section: Error bounds on the entropy of the sum of Bernoulli
random variables}

This section considers the entropy of a sum
of (possibly dependent and non-identically distributed)
Bernoulli random variables.
Section~\ref{subsection: First part of the review of
some known results} provides a review of some known results
on the Poisson approximation, via the Chen-Stein method,
that are relevant to the derivation of the new bounds (see
\cite[Section~2]{Sason}). 
Section~\ref{subsection: New error bounds on the entropy}
introduces explicit upper bounds on the error that
follows from the approximation of the entropy of a sum
of Bernoulli random variables by the entropy of a Poisson
random variable with the same mean. Some applications of
the new bounds are exemplified in Section~\ref{subsection:
Examples for the use of the mew error bounds on the entropy}.

\subsection{Background}
\label{subsection: First part of the review of some known results}

In the following, the term `distribution' refers to the 
probability mass function of an integer-valued random variable.
\begin{definition}
Let $P$ and $Q$ be two probability measures defined on a set $\mathcal{X}$.
Then, the total variation distance between $P$ and $Q$ is defined by
\begin{equation}
d_{\text{TV}}(P, Q)  \triangleq \sup_{\text{Borel} \,
A \subseteq \mathcal{X}} |P(A) - Q(A)|
\label{eq: total variation distance}
\end{equation}
where the supermum is taken w.r.t. all the Borel subsets $A$ of
$\mathcal{X}$.
If $\mathcal{X}$ is a countable set then \eqref{eq: total variation distance}
is simplified to
\begin{equation}
d_{\text{TV}}(P, Q) = \frac{1}{2} \sum_{x \in \mathcal{X}} |P(x) - Q(x)| =
\frac{||P-Q||_1}{2}
\label{eq: the L1 distance is twice the total variation distance}
\end{equation}
so the total variation distance is equal to one-half of the $L_1$-distance
between the two probability distributions.
\label{definition: total variation distance}
\end{definition}

\vspace*{0.1cm}
The following theorem combines \cite[Theorems~1 and 2]{BarbourH_1984},
and its proof relies on the Chen-Stein method:
\begin{theorem}
Let $W = \sum_{i=1}^n X_i$ be a sum of $n$ independent Bernoulli random
variables with $\expectation(X_i) = p_i$ for $i \in \{1, \ldots, n\}$,
and $\expectation(W) = \lambda$. Then, the total variation distance
between the probability distribution of $W$ and the Poisson
distribution with mean $\lambda$ satisfies
\begin{equation}
\frac{1}{32} \, \Bigl(1 \wedge \frac{1}{\lambda}\Bigr) \, \sum_{i=1}^n p_i^2
\leq d_{\text{TV}}(P_W, \text{Po}(\lambda)) \leq
\left(\frac{1-e^{-\lambda}}{\lambda}\right) \, \sum_{i=1}^n p_i^2
\label{eq: bounds on the total variation distance - Barbour and Hall 1984}
\end{equation}
where $a \wedge b \triangleq \min\{a,b\}$ for every $a, b \in \reals$.
\label{theorem: bounds on the total variation distance - Barbour and Hall 1984}
\end{theorem}

\begin{remark}
The ratio between the upper and lower bounds in
Theorem~\ref{theorem: bounds on the total variation distance - Barbour and Hall 1984}
is not larger than~32, irrespectively of the values of $\{p_i\}$. This shows that
these bounds are essentially tight. The upper bound in
\eqref{eq: bounds on the total variation distance - Barbour and Hall 1984} improves
Le Cam's inequality (see \cite{Le Cam_1960}, \cite{Steele_1994})) which states that
$ d_{\text{TV}}(P_W, \text{Po}(\lambda)) \leq \sum_{i=1}^n p_i^2 $
so the improvement, for $\lambda \gg 1$, is by the factor $\frac{1}{\lambda}$.
\end{remark}

\vspace*{0.1cm}
Theorem~\ref{theorem: bounds on the total variation distance - Barbour and Hall 1984}
provides a non-asymptotic result for the Poisson approximation of sums of independent
binary random variables via the use of the Chen-Stein method. In general, this method
enables to analyze the Poisson approximation for sums of dependent random variables. To
this end, the following notation was used in \cite{ArratiaGG_AOP} and
\cite{ArratiaGG_Tutorial90}:

Let $I$ be a countable index set, and for $\alpha \in I$, let $X_{\alpha}$ be a Bernoulli
random variable with
\newpage
\begin{equation}
p_{\alpha} \triangleq \pr(X_{\alpha}=1) = 1-\pr(X_{\alpha}=0) > 0.
\label{eq: probabilities of the Bernoulli random variables}
\end{equation}
Let
\begin{equation}
W \triangleq \sum_{\alpha \in I} X_{\alpha}, \quad \lambda \triangleq \expectation(W)
= \sum_{\alpha \in I} p_{\alpha}
\label{eq: Bernoulli sums and their mean}
\end{equation}
where it is assumed that $\lambda \in (0, \infty)$.
For every $\alpha \in I$, let $B_{\alpha}$ be a subset of $I$ that is chosen such that
$\alpha \in B_{\alpha}$. This subset is interpreted in \cite{ArratiaGG_AOP}
as the neighborhood of dependence for $\alpha$ in the sense that $X_{\alpha}$ is independent
or weakly dependent of all of the $X_{\beta}$ for $\beta \notin B_{\alpha}$. Furthermore,
the following coefficients were defined in \cite[Section~2]{ArratiaGG_AOP}:
\begin{eqnarray}
&& \hspace*{-1cm}
b_1 \triangleq \sum_{\alpha \in I} \sum_{\beta \in B_{\alpha}} p_{\alpha} p_{\beta}
\label{eq: b1} \\[0.1cm]
&& \hspace*{-1cm}
b_2 \triangleq \sum_{\alpha \in I} \sum_{\alpha \neq \beta \in B_{\alpha}} p_{\alpha, \beta},
\quad p_{\alpha, \beta} \triangleq \expectation(X_{\alpha} X_{\beta}) \label{eq: b2} \\[0.1cm]
&& \hspace*{-1cm}
b_3 \triangleq \sum_{\alpha \in I} s_{\alpha}, \quad s_{\alpha} \triangleq \expectation \bigl|
\expectation(X_{\alpha} - p_{\alpha} \, | \,
\sigma(\{X_{\beta}\})_{\beta \in I \setminus B_{\alpha}}) \bigr|  \label{eq: b3}
\end{eqnarray}
where $\sigma(\cdot)$ in the conditioning of \eqref{eq: b3} denotes the $\sigma$-algebra
that is generated by the random variables inside the parenthesis.
In the following, we cite \cite[Theorem~1]{ArratiaGG_AOP} which essentially implies
that when $b_1, b_2$ and $b_3$ are all small, then the total number $W$ of events is
approximately Poisson distributed.
\begin{theorem}
Let $W = \sum_{\alpha \in I} X_{\alpha}$ be a sum of (possibly dependent and
non-identically distributed) Bernoulli random variables $\{X_{\alpha}\}_{\alpha \in I}$.
Then, with the notation in
\eqref{eq: probabilities of the Bernoulli random variables}--\eqref{eq: b3},
the following upper bound on the total variation distance holds:
\vspace*{-0.1cm}
\begin{equation}
d_{\text{TV}}(P_W, \text{Po}(\lambda)) \leq (b_1 + b_2) \left( \frac{1-e^{-\lambda}}{\lambda}\right)
+ b_3 \Bigl(1 \wedge \frac{1.4}{\sqrt{\lambda}}\Bigr).
\label{eq: upper bound on the total variation distance by Arratia et al.}
\end{equation}
\label{theorem: upper bound on the total variation distance by Arratia et al.}
\end{theorem}

\vspace*{-0.2cm}
\begin{remark}
A comparison of the right-hand side of
\eqref{eq: upper bound on the total variation distance by Arratia et al.}
with the bound in \cite[Theorem~1]{ArratiaGG_AOP} shows a difference in
a factor of 2 between the two upper bounds. This follows from a
difference in a factor of~2 between the two definitions of the total variation
distance in \cite[Section~2]{ArratiaGG_AOP} and
Definition~\ref{definition: total variation distance} here. Note
however that Definition~\ref{definition: total variation distance} is 
consistent with, e.g., \cite{BarbourH_1984}.
\end{remark}

\begin{remark}
Theorem~\ref{theorem: upper bound on the total variation distance by Arratia et al.}
forms a generalization of the upper bound in
Theorem~\ref{theorem: bounds on the total variation distance - Barbour and Hall 1984}
by choosing $B_{\alpha} = \alpha$ for $\alpha \in I \triangleq \{1, \ldots, n\}$
(note that, due to the independence assumption of the Bernoulli random variables in
Theorem~\ref{theorem: bounds on the total variation distance - Barbour and Hall 1984},
the neighborhood of dependence of $\alpha$ is $\alpha$ itself). In this setting,
under the independence assumption,
$ b_1 = \sum_{i=1}^n p_i^2, \quad b_2 = b_3 = 0$
which therefore gives, from
\eqref{eq: upper bound on the total variation distance by Arratia et al.}, the upper
bound in 
\eqref{eq: bounds on the total variation distance - Barbour and Hall 1984}.
\label{remark: Generalization of Theorem 1 of Barbour and Hall}
\end{remark}

The following inequality holds (see
\cite[Theorem~17.3.3]{Cover_Thomas}):
\begin{theorem}
Let $P$ and $Q$ be two probability mass functions on a finite set $\mathcal{X}$
such that the $L_1$ norm of their difference is not larger than one-half, i.e.,
\begin{equation}
||P-Q||_1 \triangleq \sum_{x \in \mathcal{X}} |P(x)-Q(x)| \leq \frac{1}{2}.
\end{equation}
Then the difference between their entropies satisfies
\begin{equation}
|H(P) - H(Q)| \leq -||P-Q||_1 \; \log\left(\frac{||P-Q||_1}{|\mathcal{X}|} \right).
\label{eq: L1 bound on the entropy}
\end{equation}
\label{theorem: L1 bound on the entropy}
\end{theorem}

\newpage
The bounds on the total variation distance for the Poisson approximation
(see Theorems~\ref{theorem: bounds on the total variation distance - Barbour and Hall 1984}
and~\ref{theorem: upper bound on the total variation distance by Arratia et al.}) and
the $L_1$ bound on the entropy (see Theorem~\ref{theorem: L1 bound on the entropy})
motivate to derive a bound on $|H(W) - H(Z)|$ where $W \triangleq \sum_{\alpha \in I}
X_{\alpha}$ is a finite sum of (possibly dependent and non-identically distributed)
Bernoulli random variables, and $Z \sim \text{Po}(\lambda)$ is Poisson distributed
with mean $\lambda = \sum_{\alpha \in I} p_{\alpha}$. The problem is that the Poisson
distribution is defined on a countable set that is infinite, so the bound in
Theorem~\ref{theorem: L1 bound on the entropy} is not applicable for the considered
problem of Poisson approximation. This motivates the theorem in the next sub-section.
Before proceeding to this analysis, the following maximum entropy result of the Poisson
distribution is introduced for the special case where the Bernoulli random variables are independent. This maximum entropy result
follows directly from \cite[Theorems~7 and 8]{Harremoes_2001}.

\vspace*{0.1cm}
\begin{theorem}
The Poisson distribution $\text{Po}(\lambda)$ has the maximal entropy among all probability
distributions with mean $\lambda$ that can be obtained as sums of independent Bernoulli RVs:
\begin{eqnarray}
&& \hspace*{-0.5cm} H(\text{Po}(\lambda)) =
\sup_{S \in B_{\infty}(\lambda)} H(S) \nonumber \\
&& \hspace*{-0.5cm} B_{\infty}(\lambda) \triangleq
\bigcup_{n \in \naturals} B_n(\lambda) \\
&& \hspace*{-0.5cm} B_n(\lambda) \triangleq \left\{S: \,
S = \sum_{i=1}^n X_i, \; X_i \sim \text{Bern}(p_i), 
\; \sum_{i=1}^n p_i = \lambda \right\} \nonumber
\label{eq: maximum entropy result for the Poisson distribution}
\end{eqnarray}
where in the above sum, $\{X_i\}_{i=1}^n$ are independent Bernoulli random variables. 
Furthermore, since the supremum of the entropy over the set $B_n(\lambda)$ is monotonic
increasing in $n$, then
$$ H(\text{Po}(\lambda)) = \lim_{n \rightarrow \infty} \sup_{S \in B_n(\lambda)} H(S).$$
For $n \in \naturals$, the maximum entropy distribution in the class $B_n(\lambda)$
is the Binomial distribution of the sum of $n$ i.i.d. Bernoulli random variables  $\text{Ber}\Bigl(\frac{\lambda}{n}\Bigr)$, so
\begin{equation*}
H(\text{Po}(\lambda)) = \lim_{n \rightarrow \infty}
H\Bigl(\text{Binomial}\Bigl(n, \frac{\lambda}{n}\Bigr)\Bigr).
\end{equation*}
\label{theorem: maximum entropy result for the Poisson distribution}
\end{theorem}
\vspace*{0.1cm}

{\em Calculation of the entropy of a Poisson random variable}:
In the next sub-section we consider the approximation of the entropy of a sum
of Bernoulli random variables by the entropy of a Poisson random variable with
the same mean. To this end, it is required to evaluate the entropy of
$Z \sim \text{Po}(\lambda)$. It is straightforward to verify that
\begin{equation}
H(Z) = \lambda \log\left(\frac{e}{\lambda}\right) + \sum_{k=1}^{\infty}
\frac{\lambda^k e^{-\lambda} \log k!}{k!}
\label{eq: entropy of Poisson distribution}
\end{equation}
so the entropy of the Poisson distribution (in nats) is expressed in terms of an infinite
series that has no closed form. Sequences of simple upper and lower bounds on
this entropy, which are asymptotically tight, were derived in \cite{AdellLY_IT2010}.
In particular, for large values of $\lambda$,
\begin{equation}
H(Z) \approx \frac{1}{2} \, \log(2\pi e \lambda) -
\frac{1}{12 \lambda} - \frac{1}{24 \lambda^2}.
\label{eq: approximation of the entropy of a Poisson RV with large mean}
\end{equation}

\newpage
\subsection{New Error Bounds on the Entropy}
\label{subsection: New error bounds on the entropy}
We introduce here new error bounds on the entropy of Bernoulli sums. Due to space
limitations, the proofs are omitted. The proofs are available in the full paper version
(see \cite[Section~II.D]{Sason}).

\begin{theorem}
Let $I$ be an arbitrary finite index set with $m \triangleq |I|$. Under the
assumptions of
Theorem~\ref{theorem: upper bound on the total variation distance by Arratia et al.}
and the notation used in
Eqs.~\eqref{eq: probabilities of the Bernoulli random variables}--\eqref{eq: b3},
let
\begin{eqnarray}
&& \hspace*{-1cm} a(\lambda) \triangleq 2 \left[(b_1 + b_2) \left(\frac{1-e^{-\lambda}}{\lambda}\right)
+ b_3 \bigl(1 \wedge \frac{1.4}{\sqrt{\lambda}}\bigr) \right]
\label{eq: function a in the upper bound on the Poisson approximation of the entropy}
\\[0.2cm]
&& \hspace*{-1cm} b(\lambda) \triangleq \left[ \Bigl(\lambda \log \bigl(\frac{e}{\lambda}\bigr)\Bigr)_+ \,
+ \lambda^2 + \frac{6 \log(2\pi) + 1}{12} \right] \nonumber \\
&& \hspace*{0.5cm} \exp \left\{-\left[\lambda + (m-1) \log\left(\frac{m-1}{\lambda e} \right)
\right] \right\}
\label{eq: function b in the upper bound on the Poisson approximation of the entropy}
\end{eqnarray}
where, in \eqref{eq: function b in the upper bound on the Poisson approximation of the entropy},
$(x)_+ \triangleq \max\{x, 0\}$ for every $x \in \reals$.
Let $Z \sim \text{Po}(\lambda)$ be a Poisson random variable with mean $\lambda$.
If $a(\lambda) \leq \frac{1}{2}$ and
$\lambda \triangleq \sum_{\alpha \in I} p_{\alpha} \leq m-1$, then the difference
between the entropies (to the base~$e$) of $Z$ and $W$ satisfies the 
inequality:
\vspace*{-0.1cm}
\begin{equation}
|H(Z) - H(W)| \leq a(\lambda) \, \log\left(\frac{m+2}{a(\lambda)}\right) + b(\lambda).
\label{eq: upper bound on the Poisson approximation of the entropy}
\end{equation}
\label{theorem: upper bound on the Poisson approximation of the entropy}
\end{theorem}

\vspace*{0.1cm}
The following corollary follows from
Theorems~\ref{theorem: maximum entropy result for the Poisson distribution}
and~\ref{theorem: upper bound on the Poisson approximation of the entropy},
and Remark~\ref{remark: Generalization of Theorem 1 of Barbour and Hall}:
\begin{corollary}
Consider the setting in Theorem~\ref{theorem: upper bound on the Poisson
approximation of the entropy}, and assume that the Bernoulli random variables
$\{X_{\alpha}\}_{\alpha \in I}$ are also independent. If
$\left(\frac{1-e^{-\lambda}}{\lambda}\right)
\; \sum_{\alpha \in I} p_{\alpha}^2 \leq \frac{1}{4}$
and $\lambda \leq m-1$ then, for $Z \sim \text{Po}(\lambda)$,
\begin{eqnarray}
&& \hspace*{-1.2cm} 0 \leq H(Z) - H(W) \leq b(\lambda) + \nonumber \\
&& \hspace*{-0.8cm} 2 \left(\frac{1-e^{-\lambda}}{\lambda}\right)
\; \sum_{\alpha \in I} p_{\alpha}^2 \cdot
\log \Biggl(\frac{(m+2) \lambda}{2 (1-e^{-\lambda})
\sum_{\alpha \in I} p_{\alpha}^2} \Biggr).
\label{eq: an error bound on the Poisson approximation of the entropy
for independent Bernoulli RVs}
\end{eqnarray}
\label{corollary: upper bound on the Poisson approximation of the entropy
for independent RVs}
\end{corollary}

\vspace*{0.1cm}
The following bound forms a possible improvement of the result in
Corollary~\ref{corollary: upper bound on the Poisson approximation
of the entropy for independent RVs}. It combines the upper bound on
the total variation distance in \cite[Theorem~1]{BarbourH_1984} (see
Theorem~\ref{theorem: bounds on the total variation distance - Barbour and Hall 1984}
here) with the upper bound on the total variation distance
in \cite[Eq.~(30)]{CekanaviciusR_2006}. It is noted that the bound in
\cite[Eq.~(30)]{CekanaviciusR_2006} improves
the bound in \cite[Eq.~(10)]{Roos_2001} (see also
\cite[Eq.~(4)]{Roos_2003}).

\vspace*{0.1cm}
\begin{proposition}
Assume that the conditions in Corollary~\ref{corollary: upper bound on the
Poisson approximation of the entropy for independent RVs} are satisfied.
Then, the following inequality holds:
\begin{equation}
0 \leq H(Z) - H(W) \leq g(\underline{p}) \,
\log\left(\frac{m+2}{g(\underline{p})}\right) + b(\lambda)
\label{eq: a possibly improved error bound on the Poisson approximation
of the entropy for independent Bernoulli RVs}
\end{equation}
if $g(\underline{p}) \leq \frac{1}{2}$ and $\lambda \leq m-1$, where
\begin{eqnarray}
&& g(\underline{p}) \triangleq 2 \theta \, \min
\left\{ 1 - e^{-\lambda}, \; \frac{3}{4e (1-\sqrt{\theta})^{3/2}}
\right\} \label{eq: g} \\
&& \underline{p} \triangleq \bigl\{p_{\alpha}\bigr\}_{\alpha \in I},
\quad \lambda \triangleq \sum_{\alpha \in I} p_{\alpha}
\label{eq: lambda} \\[-0.1cm]
&& \theta \triangleq \frac{1}{\lambda} \sum_{\alpha \in I} p_{\alpha}^2.
\label{eq: theta}
\end{eqnarray}
\label{proposition: a possibly improved error bound on the Poisson approximation
of the entropy for independent Bernoulli RVs}
\end{proposition}

\newpage
\begin{remark}
From \eqref{eq: lambda} and \eqref{eq: theta}, it follows that
$$0 \leq \theta \leq \max_{\alpha \in I} p_{\alpha} \triangleq p_{\max}.$$
Furthermore, the condition $\lambda \leq m-1$ is mild since $|I|=m$ and
the probabilities $\{p_{\alpha}\}_{\alpha \in I}$ should be typically small
for the Poisson approximation to hold.
\end{remark}

\begin{remark}
Proposition~\ref{proposition: a possibly improved error bound on
the Poisson approximation of the entropy for independent Bernoulli RVs}
improves the bound in Corollary~\ref{corollary: upper bound on the
Poisson approximation of the entropy for independent RVs}
only if $\theta$ is below a certain value that depends
on $\lambda$. The maximal improvement that is
obtained by Proposition~\ref{proposition: a possibly improved
error bound on the Poisson approximation of the entropy for
independent Bernoulli RVs}, as compared to
Corollary~\ref{corollary: upper bound on the Poisson approximation
of the entropy for independent RVs}, is in the case where
$\theta \rightarrow 0$ and $\lambda \rightarrow \infty$,
and the corresponding improvement in the value of $g(\underline{p})$
is by a factor of $\frac{3}{4e} \approx 0.276$.
\end{remark}

\subsection{Some Applications of the New Error Bounds on the Entropy}
\label{subsection: Examples for the use of the mew error bounds on the entropy}

In the following, the use of
Theorem~\ref{theorem: upper bound on the Poisson approximation of the entropy}
is first exemplified when the Bernoulli random variables are independent.
It is also exemplified in a case from \cite[Section~3]{ArratiaGG_AOP}
where dependence among the Bernoulli random variables exists.
The use of Theorem~\ref{theorem: upper bound on the Poisson approximation of the entropy}
is exemplified for the calculation of error bounds on the entropy via
the Chen-Stein method.

\vspace*{0.1cm}
\begin{example}[sums of independent binary random variables] Let
$W = \sum_{i=1}^n X_i$ be a sum of $n$ independent Bernoulli random
variables where $X_i \sim \text{Bern}(p_i)$ for $i=1, \ldots, n$.
The calculation of the entropy of $W$ involves the numerical computation
of the probabilities
$$\bigl(P_W(0), P_W(1), \ldots, P_W(n)\bigr) = (1-p_1, p_1) \ast
\ldots \ast (1-p_n, p_n)$$
whose computational complexity is high for very large values of $n$,
especially if the probabilities $p_1, \ldots, p_n$ are not the same.
The bounds in
Corollary~\ref{corollary: upper bound on the Poisson approximation
of the entropy for independent RVs}
and Proposition~\ref{proposition: a possibly improved error bound on
the Poisson approximation of the entropy for independent Bernoulli RVs}
enable to get rigorous upper bounds on the accuracy of the Poisson
approximation for $H(W)$. As was explained earlier in this section,
the bound in
Proposition~\ref{proposition: a possibly improved error bound on the
Poisson approximation of the entropy for independent Bernoulli RVs}
may only improve the bound in
Corollary~\ref{corollary: upper bound on the Poisson approximation of
the entropy for independent RVs}.
Lets exemplify this in the following case:
Suppose that $$p_i = 2ai, \quad \forall \, i \in \{1, \ldots, n\},
\; a = 10^{-10}, \; n = 10^8$$ then
\begin{eqnarray}
&& \hspace*{-1cm} \lambda = \sum_{i=1}^n p_i = an(n+1) = 1,000,000.01
\approx 10^6 \, , \\[0.2cm]
&& \hspace*{-1cm} \theta = \frac{1}{\lambda} \sum_{i=1}^n p_i^2 =
\frac{2a(2n+1)}{3} = 0.0133.
\end{eqnarray}
The entropy of $Z \sim \text{Po}(\lambda)$
is $H(Z) = 8.327 \, \text{nats}$.
Corollary~\ref{corollary: upper bound on the Poisson approximation
of the entropy for independent RVs} gives that
$0 \leq H(Z) - H(W) \leq 0.588 \, \text{nats}$ and
Proposition~\ref{proposition: a possibly improved error bound on the
Poisson approximation of the entropy for independent Bernoulli RVs}
improves it to $0 \leq H(Z) - H(W) \leq 0.205 \, \text{nats}.$ Hence,
$H(W) \approx 8.224 \, \text{nats}$ with a relative error of at most
$1.2\%.$ We note that by changing the values of $a$ and $n$ to $10^{-14}$
and $10^{12}$, respectively, it follows that $H(W) \approx 12.932 \,
\text{nats}$ with a relative error of at most $0.04\%$. The enhancement
of the accuracy of the Poisson approximation in the latter case is
consistent with the law of small numbers (see, e.g., \cite{KontoyiannisHJ_2005}
and references therein).
\end{example}

\newpage
\begin{example}[random graphs]
This problem, which appears in \cite[Example~1]{ArratiaGG_AOP}, is described
as follows: On the cube $\{0,1\}^n$, assume that each of the
$n 2^{n-1}$ edges is assigned a random direction by tossing a fair
coin. Let $k \in \{0, 1, \ldots, n\}$ be fixed, and denote by
$W \triangleq W(k,n)$ the random variable that is equal to the
number of vertices at which exactly $k$ edges point outward (so
$k=0$ corresponds to the event where all $n$ edges, from a certain
vertex, point inward). Let $I$ be the set of all $2^n$ vertices,
and $X_{\alpha}$ be the indicator that vertex $\alpha \in I$ has
exactly $k$ of its edges directed outward. Then
$W = \sum_{\alpha \in I} X_{\alpha}$ with $$X_{\alpha} \sim
\text{Bern}(p), \quad p = 2^{-n} {{n}\choose{k}},
\quad \forall \alpha \in I.$$
This implies that $\lambda = {{n}\choose{k}}$ (since $|I|=2^n$).
Clearly, the neighborhood of dependence of a vertex $\alpha \in I$,
denoted by $B_{\alpha}$, is the set of vertices that are directly
connected to $\alpha$ (including $\alpha$ itself since
Theorem~\ref{theorem: upper bound on the total variation distance
by Arratia et al.} requires that $\alpha \in B_{\alpha}$). It is
noted, however, that $B_{\alpha}$ in \cite[Example~1]{ArratiaGG_AOP}
was given by $B_{\alpha} = \{\beta: \, |\beta - \alpha| = 1\}$ so
it excluded the vertex $\alpha$. From \eqref{eq: b1}, this difference
implies that $b_1$ in their example should be modified to
\begin{equation}
b_1 = 2^{-n} (n+1) {{n}\choose{k}}^2
\end{equation}
so $b_1$ is larger than its value in \cite[p.~14]{ArratiaGG_AOP}
by a factor of $1+\frac{1}{n}$ which has a negligible effect if
$n \gg 1$. As is noted in \cite[p.~14]{ArratiaGG_AOP}, if $\alpha$
and $\beta$ are two vertices that are connected by an edge,
then a conditioning on the direction of this edge gives that
$$p_{\alpha, \beta} \triangleq \expectation(X_\alpha X_\beta) =
2^{2-2n} \, {{n-1}\choose{k}} \, {{n-1}\choose{k-1}}$$
for every $\alpha \in I$ and $\beta \in B_\alpha \setminus
\{\alpha\}$, and therefore, from \eqref{eq: b2},
$$ b_2 = n \, 2^{2-n} \, {{n-1}\choose{k}} \, {{n-1}\choose{k-1}}.$$
Finally, as is noted in \cite[Example~1]{ArratiaGG_AOP}, $b_3=0$
(this is because the conditional expectation of $X_{\alpha}$
given $(X_\beta)_{\beta \in I \setminus B_{\alpha}}$ is, similarly
to the un-conditional expectation, equal to $p_{\alpha}$; i.e.,
the directions of the edges outside the neighborhood of dependence
of $\alpha$ are irrelevant to the directions of the edges connecting
the vertex $\alpha$).

In the following,
Theorem~\ref{theorem: upper bound on the Poisson approximation of the entropy}
is applied to get a rigorous error bound on the Poisson approximation of the
entropy $H(W)$.
Table~\ref{table: a random graph problem} presents numerical results for
the approximated value of $H(W)$, and an upper bound on the maximal relative
error that is associated with this approximation. Note that, by symmetry,
the cases with $W(k,n)$ and $W(n-k,n)$ are equivalent, so
$H\bigl(W(k,n)\bigr) = H\bigl(W(n-k,n)\bigr).$

\begin{table}[here!]
\caption{Numerical results for the Poisson approximations of the entropy $H(W)$
($W = W(k,n)$) by the entropy $H(Z)$ where $Z \sim \text{Po}(\lambda)$, jointly with
the associated error bounds of these approximations. These error bounds are calculated from
Theorem~\ref{theorem: upper bound on the Poisson approximation of the entropy} for
the random graph problem in Example~\ref{example: a random graph problem}.} \vspace*{0.3cm}
\label{table: a random graph problem}
\centering
\renewcommand{\arraystretch}{1.5}
\begin{tabular}{|c|c|c|c|c|} \hline
$n$ & $k$  &  $ \lambda = {{n}\choose{k}} $  & $H(W) \approx$ & Maximal relative error \\  \hline
30 & 27 & $4.060 \cdot 10^3$ & 5.573 \text{nats} & 0.16\% \\
30 & 26 & $2.741 \cdot 10^4$ & 6.528 \text{nats} & 0.94\% \\
30 & 25 & $1.425 \cdot 10^5$ & 7.353 \text{nats} & 4.33\% \\ \hline
50 & 48 & $1.225 \cdot 10^3$ & 4.974 \text{nats} & $1.5 \cdot 10^{-9}$\\
50 & 44 & $1.589 \cdot 10^7$ & 9.710 \text{nats} & $1.0 \cdot 10^{-5}$ \\
50 & 40 & $1.027 \cdot 10^{10}$ & 12.945 \text{nats} & $4.8 \cdot 10^{-3}$ \\ \hline
100 & 95 & $7.529 \cdot 10^7$ & 10.487 \text{nats} & $1.6 \cdot 10^{-19}$ \\
100 & 85 & $2.533 \cdot 10^{17}$ & 21.456 \text{nats} & $2.6 \cdot 10^{-10}$ \\
100 & 75 & $2.425 \cdot 10^{23}$ & 28.342 \text{nats} & $1.9 \cdot 10^{-4}$ \\
100 & 70 & $2.937 \cdot 10^{25}$ & 30.740 \text{nats} & $2.1\%$ \\ \hline
\end{tabular}
\end{table}
\label{example: a random graph problem}
\end{example}

\subsection{Generalization: Bounds on the Entropy for a Sum of Non-Negative, Integer-Valued and Bounded Random Variables}
\label{subsection: generalization of the bounds on the entropy for the sum of integer-valued random variables}
We introduce in \cite[Section~II-E]{Sason} a generalization of the bounds in
Section~\ref{subsection: New error bounds on the entropy} that considers the
accuracy of the Poisson approximation for the entropy of a sum of non-negative,
integer-valued and bounded random variables.
\newpage
This generalization is enabled via
the combination of the proof of
Theorem~\ref{theorem: upper bound on the Poisson approximation of the entropy}
for sums of Bernoulli random variables with the
approach of Serfling in \cite[Section~7]{Serfling_1978}.

\end{document}